\documentclass[english,prd,twocolumn,superscriptaddress,nofootinbib,preprintnumbers]{revtex4}
\usepackage[latin1]{inputenc}
\usepackage{graphicx}
\usepackage{amssymb}
\usepackage{color}
\usepackage{float}
\usepackage{amsmath}
\usepackage{amsfonts}
\usepackage{dcolumn}
\usepackage{hyperref}
\usepackage{subfigure}

\def\beq{\begin{equation}}
\def\eeq{\end{equation}}
\def\bea{\begin{eqnarray}}
\def\eea{\end{eqnarray}}

\begin{document}

\title{Early-time cosmological solutions in Einstein-scalar-Gauss-Bonnet theory}

\author{Panagiota Kanti}
\affiliation{Department of Physics,University of Ioannina, Ioannina GR-45110, Greece}
\affiliation{Department of Physics, National and Kapodistrian University of Athens, Athens, Greece}

\author{Radouane Gannouji}
\affiliation{Instituto de F\'{\i}sica, Pontificia Universidad  Cat\'olica de Valpara\'{\i}so, 
Casilla 4950,\\ Valpara\'{\i}so, Chile}

\author{Naresh Dadhich}
\affiliation{Centre for Theoretical Physics, Jamia Millia Islamia, New Delhi-110025, India} 
\affiliation{Inter-University Centre for Astronomy \& Astrophysics, Post Bag 4, Pune 411 007, India}

\begin{abstract}
In this work, we consider a generalised gravitational theory that contains the Einstein term,
a scalar field and the quadratic Gauss-Bonnet term. We focus on the early-universe
dynamics, and demonstrate that a simple choice of the coupling function between the
scalar field and the Gauss-Bonnet term and a simplifying
assumption regarding the role of the Ricci scalar can lead to new, analytical, elegant
solutions with interesting characteristics. We first argue, and demonstrate in the context
of two different models, that the presence of the Ricci scalar in the theory at early times,
when the curvature is strong, does not affect the actual cosmological solutions. By
considering therefore a pure scalar-GB theory with a quadratic coupling function 
we derive a plethora of interesting, analytic solutions: for a negative coupling parameter, we
obtain inflationary, de Sitter-type solutions or expanding solutions with a de Sitter
phase in their past and a natural exit mechanism at later times; for a positive coupling
function, we find instead singularity-free solutions with no Big-Bang singularity.
We show that the aforementioned solutions arise only for this particular choice of coupling
function, a result that may hint to some fundamental role that this coupling function
may hold in the context of an ultimate theory. 
\end{abstract}

\maketitle

\section{Introduction}

In the quest for the final theory, that would unify the gravitational interactions with
their particle physics analogs and describe them correctly at arbitrarily large energy
scales, the generalised gravitational theories have played a primary role. In their majority,
these theories are purely geometrical in nature and involve extra dimensions: well-known
examples are superstring theory \cite{strings}, the Lovelock theory \cite{Lovelock} or
the novel theories with extra dimensions \cite{ADD, RS}. For a 4-dimensional observer,
however, the dynamics and content of this higher-dimensional, fundamental theory
is translated into the appearance of new terms, describing gravity or a number of
additional fields, in the context of the 4-dimensional effective theory. Inspired by
the aforementioned theories, many variants of modified gravitational theories have
been constructed over the years, and their implications for gravity and cosmology
were extensively studied. 

The most usual way to modify the gravitational interactions in a 4-dimensional
context is via the addition of gravitational terms that involve higher powers of
curvature. In the context of the heterotic superstring effective theory
\cite{Zwiebach, Gross, Tseytlin}, for
instance, the Einstein term is supplemented by quadratic curvature terms, such
as the Gauss-Bonnet term $R^2_{GB}$ or the $R \tilde R$ term with the latter,
however, being trivially zero for a Friedmann-Lema\^itre-Robertson-Walker background. 
The Gauss-Bonnet term is also the first-order correction to the Einstein term
in the Lovelock theory, that is considered as a natural generalisation of Einstein's
theory of gravity in a higher number of dimensions \cite{Dadhich}. In 4 dimensions,
however, the Gauss-Bonnet (GB) term is a topological invariant and adds nothing to
the field equations of the theory unless it is coupled to an additional field. 
Inspired by superstring theory, the usual way is to couple the GB term to a
scalar field. By doing that, the GB term remains in the theory and has been
shown to lead to a variety of new solutions: singularity-free cosmological
solutions \cite{Antoniadis, KRT}, novel hairy black holes \cite{KMRTW, Torii} or even
traversable wormholes \cite{KKK} -- solutions that were absent in the traditional
General Relativity that contains only the Ricci scalar term (for a review on this
type of generalised gravitational theories and further references, see \cite{Nojiri}). 
Also, since the field equations remain second order for the Lovelock
theory in general and GB in particular, they are widely known to be ghost-free
theories contrary to some brane models like DGP (see e.g. \cite{Gregory:2008bf}
for a summary) or higher-derivative theories, which often suffer from the Ostrogradski
ghost \cite{Chen:2012au} (however, see \cite{Gleyzes:2014dya} for a counterexample).
These theories have also been intensively studied in various realistic contexts: by
considering solar system constraints \cite{Amendola:2007ni}, as a dark energy model
with CMB and galaxy distribution constraints \cite{Koivisto:2006xf} or as an inflationary model \cite{Carter:2005fu,Leith:2007bu}.

In this work, we will focus on the dynamics of early universe and look for the
corresponding cosmological solutions. During the early universe, it is natural to
expect that a string-inspired theory would describe better its dynamics -- that
is why here we consider a 4-dimensional theory containing the Einstein term and
a scalar field (one of the many of string theory) coupled non-minimally to gravity
through a  general coupling function to the quadratic Gauss-Bonnet term. 
However, we will go one step beyond that: we will demonstrate, in the context of
two different models, that the presence of the Ricci scalar adds nothing to the
dynamics of the universe at early times where the curvature is strong; in fact,
it is the coupled system of scalar field and GB term that dominates and 
governs the evolution of the universe. Motivated by this, we will then look for and
derive, exclusively via analytical means, cosmological solutions with interesting
characteristics in the context of the pure scalar-GB
theory. For the particular case of a quadratic coupling function between 
the scalar field and the GB term, we will present two classes of solutions that
are distinguished by the sign of the coupling function constant: for a negative
coupling constant, the system of equations supports cosmological solutions that
are either purely de Sitter-like or have a de Sitter inflationary phase in their
past and a natural exit mechanism at later times; for a positive coupling 
constant, a class of singularity-free solutions emerges instead.

Although these solutions are derived for a particular form of the coupling function,
it has been shown in the past that a polynomial, even coupling function, such as
the quadratic one chosen here, shares many common characteristics with the
actual coupling function between the moduli fields and the GB term in the heterotic
superstring effective theory \cite{KRT}.  Therefore, our present analysis, that proposes a new
approach in deriving early-time cosmological solutions, will be of relevance to
the superstring effective theory itself, but also to any other generalised, string-inspired
theory that contains the GB term coupled to a scalar field. All these theories are
bound to have in the phase-space of their solutions, the aforementioned classes
of solutions as early-time asymptotics. In addition, the analytical derivation
of the solutions, in contrast to the usual numerical means employed systematically
when the Ricci term is kept in the theory, allows for a more comprehensive study
of the properties of the found solutions.   
The presence of the Ricci scalar inevitably becomes important, as the universe expands,
and a transition between the scalar-GB-dominated phase and a subsequent one that
prepares the universe for a traditional Ricci-scalar-driven cosmology will eventually
take place. Such a transitory phase is studied at the final part of this work.

Our approach is very similar to the two important conjectures about asymptotic dynamics
at early times proposed by Belinsky, Khalatnikov and Lifshitz (BKL). The two conjectures state
that matter content \cite{Lifshitz:1963ps,Belinsky:1970ew} and spatial derivatives \cite{Belinskii:1972sg}
are not dynamically significant near the initial singularity. With the same spirit, we will conjecture
and prove in a class of models that lower-curvature terms in the Riemann tensor are negligible
near the singularity compared to higher-curvature terms.

The outline of our manuscript is as follows: in Section 2, we present the theoretical
framework of our analysis and the derived set of field equations. In Section 3, through
the use of a toy model, we demonstrate how the presence of the Ricci term in the theory
merely complicates the analysis without changing anything in the actual dynamics of
the solution. We reinforce this argument in Section 4, where the complete theory of
the Ricci scalar, a scalar field and the Gauss-Bonnet term is studied for the case of a
linear coupling function. In Section 5, we consider the pure scalar-GB theory with
a quadratic coupling function, and derive a variety of early-time cosmological
solutions with attractive characteristics. In Section 6, we briefly consider the case
of a general polynomial coupling function, and in Section 7, we study an indicative
transitory phase for the universe as it passes from a scalar-GB-dominated phase
to one where the Ricci scalar starts being important. We present our conclusions in
Section 8. 


\section{The Theoretical Framework}

In this work, we consider a string-inspired gravitational theory that contains,
apart from the Ricci scalar, a scalar field $\phi$ coupled non-minimally to gravity
via a general coupling function $f(\phi)$ to the quadratic Gauss-Bonnet term.
Such a theory is described by the following action functional
\begin{eqnarray}  
{\cal S}=\int d^4x \sqrt{-g} \left[\frac{R}{2}-\frac{(\nabla \phi)^2}{2} +
\frac{1}{8}\,f(\phi) R^2_{\rm GB}\right],
\label{action}
\end{eqnarray} 
where the Gauss-Bonnet term $R^2_{\rm GB}$ is defined as
\beq
R^2_{\rm GB} = R_{\mu\nu\rho\sigma} R^{\mu\nu\rho\sigma}
- 4 R_{\mu\nu} R^{\mu\nu} + R^2\,.
\eeq
Since the focus of this work will be the dynamics of the universe at very early times
where high-energy and strong-curvature effects are expected to be the dominant ones,
throughout this work, we will assume that any additional distribution of matter or
energy, apart from the scalar field, plays only a secondary role and thus will be ignored. 

The variation of the action (\ref{action}) with respect to the scalar field $\phi$ and the
metric tensor $g_{\mu\nu}$ leads to the scalar and gravitational field equations, respectively;
these have the form
\beq
\frac{1}{\sqrt{-g}}\,\partial_\mu\left[\sqrt{-g}\,\partial^\mu \phi \right] +
\frac{1}{8}\,f' R^2_{GB}=0\,,
\label{dilaton-eq-cov}
\eeq
and
\beq
R_{\mu\nu}-\frac{1}{2}g_{\mu\nu}R + P_{\mu \alpha \nu \beta}\nabla^{\alpha \beta}f
=\partial_\mu\phi \partial_\nu \phi
- g_{\mu \nu}\,\frac{(\nabla\phi)^2}{2}\,,
\label{Einstein-eqs}
\eeq
where $f' \equiv df/d\phi$, and $P_{\mu\alpha\nu\beta}$ is defined as 
\beq
P_{\mu \alpha \nu \beta}=R_{\mu \alpha \nu \beta}+2g_{\mu[\beta}R_{\nu]\alpha}+
2g_{\alpha[\nu}R_{\beta]\mu}+Rg_{\mu[\nu}g_{\beta]\alpha}\,.
\eeq
We note that if, the Gauss-Bonnet term is altogether ignored in the theory, the scalar field
looses its potential; on the other hand, if the scalar coupling function is a constant,
then $\nabla^{\alpha \beta}f=0$ and the contribution of the Gauss-Bonnet term to the
gravitational field equations (\ref{Einstein-eqs}) vanishes. We may therefore conclude
that $\phi$ and $R^2_{GB}$ seem to form an independent pair of quantities that mutually
support each other. Recall that it is the metric that acts as potential in General relativity
and it is coupled to the velocity, $u^a$, in the Lagrangian, $g_{ab}u^au^b$, for particle
motion. It should be noted that the coupling of the scalar field  with the Gauss-Bonnet
term is in the same general relativistic form and spirit. It is therefore a very desirable 
feature of the theory. 

We will also assume that the line-element has the Friedmann-Lema\^itre-Robertson-Walker form
\beq
ds^2 = -dt^2 +a^2(t)\left[\frac{dr^2}{1-kr^2}+r^2\,(d\theta^2 + 
\sin^2\theta\,d\varphi^2)\right],
\label{metric}
\eeq
that describes a homogeneous and isotropic universe with a scale factor $a(t)$ and
spatial curvature $k=0, \pm 1$. For the above metric ansatz, the Gauss-Bonnet term
takes the explicit form
\beq
R^2_{GB}=24\,\bigl(H^2+\frac{k}{a^2}\bigr) (H^2+\dot H)\,,
\label{GB}
\eeq
where $H \equiv \dot a/a$ is the Hubble parameter and the dot denotes
derivative with respect to time.
Using the above geometrical quantities, the scalar (\ref{dilaton-eq-cov}) and gravitational
field equations (\ref{Einstein-eqs}) reduce to the following system of three, ordinary but
coupled, differential equations 
\begin{align}
\ddot \phi+3 H \dot \phi-3f' \bigl(H^2+\frac{k}{a^2}\bigr)
(H^2+\dot H) &=0\,, \label{dilaton-eq-0} \\
3(1+H \dot f)\bigl(H^2 +\frac{k}{a^2}\bigr) &= \frac{\dot \phi^2}{2}\,,
\label{Ein-1}\\
2(1+H\dot f) (H^2+\dot H)+(1+\ddot f) \bigl(H^2+
\frac{k}{a^2}\bigr) &=-\frac{\dot \phi^2}{2}\,.\label{Ein-2}
\end{align}

In the following sections, we will look for cosmological solutions supported by the above
set of equations both in the presence and in the absence of the Ricci scalar. We will start
with a simple toy model, that will demonstrate the role of the Ricci scalar in the context
of the complete theory; the same task will then be performed in the framework of a less
restricted model. Using the derived insight of when the Ricci scalar may be ignored from
the theory, we will then search for solutions where the Ricci scalar is negligible compared to
or of the same order as the Gauss-Bonnet term.


\section{A Toy Model}

In this section, we will look for solutions of the set of Eqs. (\ref{dilaton-eq-0})-(\ref{Ein-2})
on which we impose the constraint of the vanishing of the Gauss-Bonnet term, $R^2_{GB}=0$.
The $\phi=const.$ is then a solution of the scalar field equation which is trivially satisfied.
But, the Einstein's equations, since now $\dot\phi=\ddot\phi=0$, demand that the following
two constraints
\beq
\frac{k+{\dot a}^2}{a^2}=0\,, \qquad {\rm and} \qquad \ddot a=0
\eeq
should be {\it simultaneously} satisfied. As it was shown in \cite{KRT}, for $k=0,+1$, the
above equations lead to a static universe with an arbitrary or infinite radius,
respectively, while for $k=-1$ we obtain a linearly expanding universe with 
$a(t)=A t+B$ and a singularity at a finite time.

We will thus search for solutions with $\dot \phi \neq 0$. Since $R^2_{GB}=0$, the scalar
equation is easily integrated once to give 
\beq
\dot \phi(t)=\frac{C}{a^3(t)}\,, \label{phi-dot-0}
\eeq
where $C$ an arbitrary integration constant. The solution for the scale factor is most
easily given by the constraint $R^2_{GB}=0$ itself, or equivalently
\beq
\Bigl(\frac{k+{\dot a}^2}{a^2}\Bigr) \ddot a=0\,,
\label{GB-choices}
\eeq
where the two multiplying factors could be {\it independently} zero or not.
In fact, it is only for the second choice, $\ddot a=0$, that a non-trivial solution
for the scalar field is allowed. Then, we easily find that $a(t)=A t +B$ again,
but now this solution holds for all values of $k$. Taking the sum of
Eqs. (\ref{Ein-1})-(\ref{Ein-2}), we find the constraint
\beq
\ddot f +3\dot f\,\frac{\dot a}{a}+4=0\,,
\eeq
with solution
\beq
f(t)=-\frac{c_1}{2A (A t+B)^2} -\frac{Bt}{A}-\frac{t^2}{2} +c_2\,,
\eeq
where $c_{1,2}$ are again integration constants. The scalar field itself can easily be
found from Eq. (\ref{phi-dot-0}) to have the form
\beq
\phi(t)=-\frac{C}{2A (At+B)^2} +D\,. \label{phi-sol-0}
\eeq
From a field-theory point-of-view, it would be much preferable to
express the coupling function $f$ in terms of the field $\phi$ instead of the
time coordinate. Comparing the above two equations, and upon a convenient
choice of the integration parameter $D$, i.e. $D=0$, we see that we can write
\beq
f(\phi)= f_1\,\phi+ \frac{f_2}{\phi} +f_3\,, \label{f-Ricci}
\eeq
where $f_i$ are constants given in terms of $(A,B,C,c_1,c_2)$.

Let us now ignore the presence of the Ricci scalar in the theory. Since the scalar-field
equation (\ref{dilaton-eq-0}) and the constraint (\ref{GB-choices}) remain unaltered, both
the solution for the scalar field (\ref{phi-sol-0}) and the linearly-expanding form of
the scale factor are still valid. However, the Einstein field equations now take the
simplified from
\beq
\frac{3\,(k+{\dot a}^2)}{a^2}\,\dot f\frac{\dot a}{a}-\frac{{\dot \phi}^2}{2}=0\,,
\label{Ein-1-new}
\eeq
%
\beq
\frac{(k+{\dot a}^2)}{a^2}\,\ddot f+\frac{2 \ddot a \dot f}{a}\,\frac{\dot a}{a}+
\frac{{\dot \phi}^2}{2}=0\,,
\label{Ein-2-new}
\eeq
which, when combined with each other, lead to the constraint 
\beq
\ddot f +3\dot f\,\frac{\dot a}{a}=0\,,
\label{f-noRicci}
\eeq
for the coupling function. The above can be easily integrated and expressed in terms
of the scalar field to obtain
\beq
f(\phi)= f_0\,\phi+ f_3\,.
\eeq
Therefore, assuming that the Ricci scalar can be ignored, we obtain the same solutions
for the scalar field and scale factor as in the presence of it, with the only
change appearing in the expression of the coupling function that now takes a simpler form.
Looking more carefully, the two expressions (\ref{f-Ricci}) and (\ref{f-noRicci}) are
equivalent in the limit where the scalar field takes very large values; from Eq. (\ref{phi-sol-0}),
this happens as we approach the initial singularity, $(A t+B) \rightarrow 0$. Therefore,
as expected, in the presence of the quadratic Gauss-Bonnet term in the theory the Ricci
scalar adds nothing to the dynamics of the universe at the very early-time limit. 


\section{The Complete Theory with a Linear Coupling}

In this section, we will attempt to reinforce the conclusion drawn in the previous
section regarding the role of the Ricci scalar in the early-universe cosmology but
in the context of a more realistic set-up. We will therefore look for 
physically-interesting solutions following from the complete set of Eqs.
(\ref{dilaton-eq-0})-(\ref{Ein-2}) under the assumptions that the Gauss-Bonnet
term is not zero and that the function $f(\phi)$ is a non-trivial function of the 
field $\phi$. Then, using the relations 
\beq
\dot f=f'\,\dot \phi\,, \qquad 
\ddot f=f''\,{\dot \phi}^2 + f'\,\ddot \phi\,,
\label{fdot}
\eeq
and adding Eqs. (\ref{Ein-1}) and (\ref{Ein-2}), we end up with the
constraint
\begin{align}
\Bigl(H^2 +\frac{k}{a^2}\Bigr)\Bigl[4 &+f'(\ddot\phi + 3H{\dot \phi}) + 
f'' {\dot \phi}^2\Bigr]\nonumber\\
&+2(H^2 +\dot H)(1+f' H{\dot \phi})=0\,.
\label{Eq1and2-R}
\end{align}
We may now use Eqs. (\ref{dilaton-eq-0}) and (\ref{Ein-1}) to replace the
combination $(\ddot\phi+ 3H{\dot \phi})$ and $\dot \phi^2$ in the second and
third term, respectively, of the above equation. Then, we arrive at
\begin{align}
4\Bigl(H^2 +\frac{k}{a^2}\Bigr) &+ \left[2(H^2 +\dot H) + 
6f''\Bigl(H^2 +\frac{k}{a^2}\Bigr)^2\right](1+f'H\dot \phi) \nonumber\\
&+ 3f'^2 \Bigl(H^2 +\frac{k}{a^2}\Bigr)^2 (H^2 +\dot H)=0\,.
\label{Eq1and2-R-1}
\end{align}

In this work, we will consider a polynomial form for the scalar coupling function,
i.e. $f(\phi)=\lambda\,\phi^n$, where $\lambda$ a constant and $n$ an integer.
The case with $n=0$ results into a constant coupling function and is equivalent to
ignoring the Gauss-Bonnet term from the theory. The particular case of $n=1$
will be studied in this section while the case with $n=2$ will be considered in Section 5.

Let us therefore focus on the scalar equation (\ref{dilaton-eq-0}) and use that
$f(\phi)=\lambda \phi$; then, it can be brought to the form
\beq
\frac{d(\dot \phi\,a^3)}{dt}=3\lambda \ddot a\,(k+\dot a^2)\,,
\eeq
which can be straightforwardly integrated to yield the relation
\beq
\dot \phi=\frac{C}{a^3} + \frac{\lambda \dot a\,(3k+\dot a^2)}{a^3}\,,
\label{rel-phi-adot}
\eeq
with $C$ an integration constant. The second term in the above expression, proportional
to the coupling constant $\lambda$, is clearly the one associated to the Gauss-Bonnet
term, while the first one appeared also in Eq. (\ref{phi-dot-0}) in the absence of a
potential. As we are now interested in studying the properties of solutions arising in the
presence of the quadratic GB term, we will set for simplicity $C=0$, and make a comment
on the role of a non-vanishing value of $C$ later in this section.

\begin{figure}
\includegraphics[width = 0.45 \textwidth] {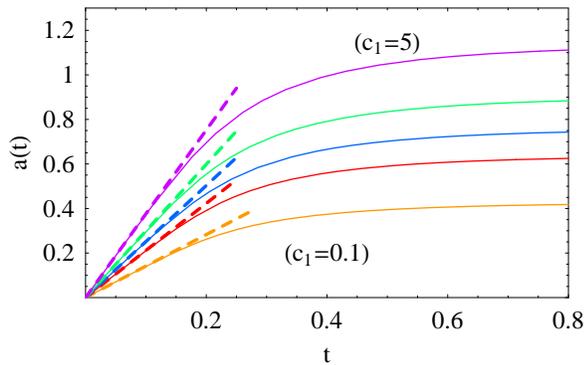}
    \caption{The scale factor $a(t)$ in terms of time, for a linear coupling function and
$\lambda=0.1$, and for the values $c_1=0.1, 0.5,1,2,5$ (from bottom to top).}
   \label{Linear}
\end{figure}

For the case of linear coupling, Eq. (\ref{Eq1and2-R-1}) is also simplified since $f''=0$.
Moreover, we may use Eq. (\ref{Ein-1}) to replace the combination $(1+f'H \dot \phi)$
in terms of $\dot\phi^2$. Then, Eq. (\ref{Eq1and2-R-1}) takes the simpler, more
explicit form
\beq
3(k+\dot a^2)^2 \left[4 +\lambda^2\,\frac{3 \ddot a}{a^3}\,(k+\dot a^2)\right]
+\ddot a \,a^3 \dot\phi^2=0.
\eeq
Now, keeping only the second term in Eq. (\ref{rel-phi-adot}) and using this
relation to substitute $\dot\phi$ in the equation above, the latter constraint is
finally rewritten as
\beq
3(k+\dot a^2)^2\left[4 +\lambda^2\,\frac{3 \ddot a}{a^3}\,(k+\dot a^2)\right]
+\lambda^2\,\frac{\ddot a {\dot a}^2}{a^3}\,(3 k +\dot a^2)^2=0.
\eeq
The above equation does not involve $\phi$ anymore, and it can be integrated once to give
\beq
3a^4=-\lambda^2 \left(\frac{13 k \dot a^2}{2} +\frac{5 \dot a^4}{2}
+\frac{2 k^3}{k+\dot a^2}\right) + c_1\,,
\label{adot-linear}
\eeq
with $c_1$ an integration constant again. Unfortunately, for $k \neq 0$, the above 
cannot be easily solved for $\dot a$ to yield, via another integration, the form of
the scale factor. However, for the case of a flat universe ($k=0$), we easily write that
\beq
\dot a ^4=\frac{2c_1}{5\lambda^2}\,\left(1-\frac{3 a^4}{c_1}\right)
\label{dota-C0}
\eeq
with solution 
\beq
a(t)\,F\left[\frac{1}{4},\frac{1}{4},\frac{5}{4};\frac{3 a^4(t)}{c_1}\right]=
\left(\frac{2c_1}{5\lambda^2}\right)^{1/4}(t+t_0)\,.
\label{solan1C0}
\eeq
In the above, $F(a,b,c;x)$ stands for the hypergeometric function whose convergence
demands that $a(t) \leq a_{max}=(c_1/3)^{1/4}$ -- from Eq. (\ref{adot-linear}) and
for $k=0$, it follows that $c_1$ must be indeed a positive constant for a real solution
for the scale factor $a(t)$ to exist. As a result, the scale factor is bounded from
above with the constant $c_1$ determining its upper value. On the other hand,
in the limit $a \rightarrow 0$, the hypergeometric function goes to unity,
and the relation between the scale factor and the time coordinate
becomes linear; this of course signals the existence of a singularity at a
finite value of the time-coordinate. Figure \ref{Linear} depicts the profile of the scale
factor as it starts from the initial singularity, follows an increasing phase and
eventually reaches its maximum value determined by $c_1$. The value of the
coupling parameter $\lambda$ affects the slope of the uprising curve at early times.

Let us make, at this point, a comment on the role of a non-vanishing value of the
parameter $C$. If we keep both terms in Eq. (\ref{rel-phi-adot}) and follow a similar
analysis as above, then, for $k=0$ again, we obtain, instead of Eq. (\ref{dota-C0}),
the following equation
\beq
3a^4=\frac{C^2}{2 \dot a^2} -2 \lambda C \dot a-
\frac{5 \lambda^2}{2}\,\dot a^4 +c_1\,.
\label{sol-a-Cneq0}
\eeq
The above is again difficult to solve for $\dot a$ and integrate further, however, we
may study particular limits. For a non-vanishing value of $C$, ignoring the first two
terms, as we did in the analysis above, would still be justified in a regime where
$\dot a$ takes very large values. Drawing experience from the known cosmological
solutions, we know that the rate of expansion of the universe is the largest close to
the initial singularity; as we are indeed interested in the early-universe dynamics,
the role of $C$ is therefore negligible and our assumption was justified.\footnote{For
completeness, let us note that in a regime where $\dot a$ would be very small
and thus the first term on the right-hand-side of Eq. (\ref{sol-a-Cneq0}) would
dominate instead, the corresponding solution for the scale factor is given by the
form $a(t) \sim (A t+B)^{1/3}$. For intermediate values of $\dot a$,  where the
second term in Eq. (\ref{sol-a-Cneq0}) would dominate, the scale factor behaves
as $a(t) \sim (t+t_0)^{-1/3}$.}

Looking for further support to our argument that the presence of the Ricci scalar
adds very little to the dynamics of the early-time cosmological solutions, we will
now adopt a different approach: we will focus on the early-time limit, and ignore
from the beginning the Ricci scalar from the field equations. In the absence of $R$,
the gravitational equation (\ref{Ein-1-new}) can be straightforwardly solved for
$\dot \phi$ to yield
\beq
\dot \phi=6f'H \Bigl(H^2+\frac{k}{a^2}\Bigr)\,,
\label{phi-dot}
\eeq
even for a general coupling function $f(\phi)$ and without any need for integration.
If we then follow a similar analysis as at the beginning of this section, i.e. take
the sum of Eqs. (\ref{Ein-1-new}) and (\ref{Ein-2-new}), and use the scalar equation
(\ref{dilaton-eq-0}) and Eq. (\ref{phi-dot}) in there, we obtain the new constraint
\begin{align}
3f'^2\Bigl(H^2+\frac{k}{a^2}\Bigr)\Bigl[ \Bigl(5&H^2 +\frac{k}{a^2}\Bigr)\,(\dot H +H^2)\nonumber\\
&+ 12 f'' H^2 \Bigl(H^2+\frac{k}{a^2}\Bigr)^2\Bigr]=0\,.
\label{Eq1and2-v1}
\end{align}
The quantity $f'$ is not allowed to be zero since then the Gauss-Bonnet term would
be eliminated from the theory. The same holds for the combination $(H^2 +k/a^2)$
since this term is part of both the Gauss-Bonnet expression and $\dot \phi$,
Eqs. (\ref{GB}) and (\ref{phi-dot}) respectively. Therefore, it is the expression inside
the square brackets that should vanish instead; for a general polynomial form of
the coupling fucntion, $f(\phi)=\lambda\,\phi^n$, this may be re-written as
\beq
(k+5{\dot a}^2)\,\ddot a +
\frac{12 {\dot a}^2}{a^3}\,(k+{\dot a}^2)^2\,\lambda n (n-1)\,\phi^{n-2}=0\,.
\label{Eq1and2-v2}
\eeq
Therefore, for the linear case with $n=1$, the second term in the above
equation vanishes which leaves us with a much simpler constraint: its
general solution is a linear function of time, i.e. $a(t)=A t+B$.
This is in accordance to the behaviour obtained in the first part of this section
when the early-time limit of the complete solution (\ref{solan1C0}) was considered.
The dashed lines in Fig. \ref{Linear} represent the linear-in-time solution for the
scale factor, derived above in the absence of the Ricci scalar in the theory - the
agreement with the early-time behaviour of the complete solution is more than evident. 
Therefore, we conclude again that, by ignoring the presence of the Ricci scalar, 
we significantly simplify the analysis and still obtain exactly the same early-time
cosmological solution.

As we anticipate from the above, the solution for the scalar field may be derived,
in the same early-time limit, from either Eq. (\ref{rel-phi-adot}) with $C=0$ or
from its simpler analog (\ref{phi-dot}): in both cases, we easily obtain that at
early times 
\beq
\phi(t)=\phi_0 -\frac{3 \lambda\,(k+A^2)}{(A t +B)^2}\,,
\eeq
where $A$, $B$, and $\phi_0$ are arbitrary integration constants. The above describes a
decaying scalar field as the universe expands with a singular behaviour at the initial singularity.


\section{The Scalar-GB Theory with Quadratic Coupling}

In this section, we address the case of the quadratic coupling function of the
scalar field to the Gauss-Bonnet term, $f=\lambda \phi^2$. The constraint
(\ref{Eq1and2-R-1}), derived in the context of the complete 
Einstein-scalar-GB theory, is valid for an arbitrary $f(\phi)$, therefore it could be
applied for this case, too. However, the scalar equation (\ref{dilaton-eq-0})
now cannot be easily integrated, and as a result the set of field equations
cannot be decoupled. Despite our persistent efforts, no way forward could be
found via analytical calculations.

Nevertheless, if one is interested strictly in the early-universe dynamics, the results
of Sections 3 and 4 point towards ignoring the Ricci term from the very beginning in
order to simplify the analysis and increase the chances of deriving viable
cosmological solutions via analytical means. Equation (\ref{Eq1and2-v2}), valid in
the context of the pure scalar-Gauss-Bonnet theory, has been derived for a general
polynomial coupling function and it can be applied directly for the quadratic case
with $n=2$. In that case, the $\phi$-dependence disappears and the constraint
becomes a differential equation only for the scale factor $a(t)$. It can be conveniently
separated as follows
\beq
\frac{(k+5{\dot a}^2)\,\ddot a}{\dot a (k+{\dot a}^2)^2}=
-\frac{24 \lambda \dot a}{a^3}\,,
\eeq
leading eventually, after integrating both sides with respect to time, to the relation
\beq
\frac{12 \lambda}{a^2}=-\frac{1}{k}\,\ln\left(\frac{\sqrt{k+{\dot a}^2}}{\dot a}\right)
-\frac{2}{(k+{\dot a}^2)} +C_1\,.
\eeq
In the above, $C_1$ is an arbitrary integration constant. For $k \neq 0$, the above
relation is of a transcendental form and thus impossible to solve for $\dot a$. We
are thus forced to consider again the case of a flat universe, in which case we
obtain the simple differential equation
\beq
\frac{5}{2 {\dot a}^2}=C_1 -\frac{12 \lambda}{a^2}\,.
\label{general}
\eeq
In fact, a variety of cosmological solutions with interesting characteristics may be
derived from the above simple equation depending on the values of the integration
constant $C_1$ and the Gauss-Bonnet coupling parameter $\lambda$. Below, we
present a comprehensive analysis of all the ensuing solutions. 


\subsection{The case with $C_1=0$}

\begin{figure}[t]
\includegraphics[width = 0.45 \textwidth] {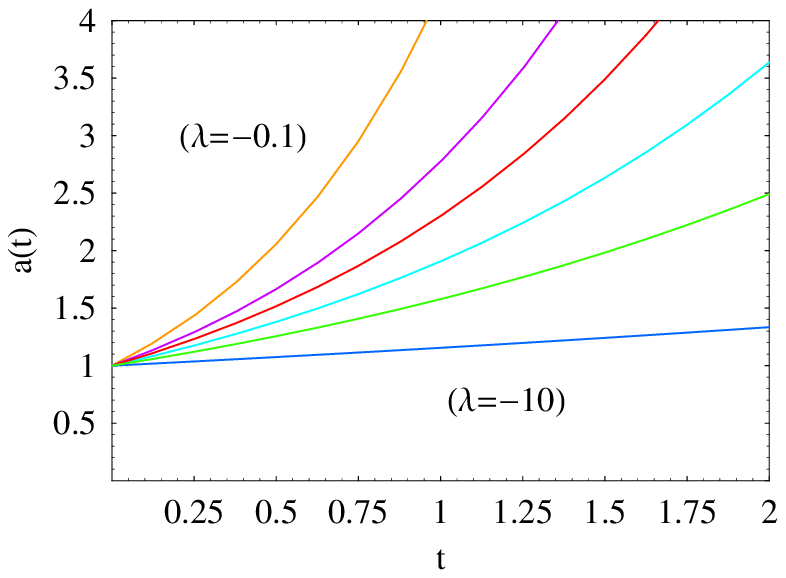}
\includegraphics[width = 0.45 \textwidth] {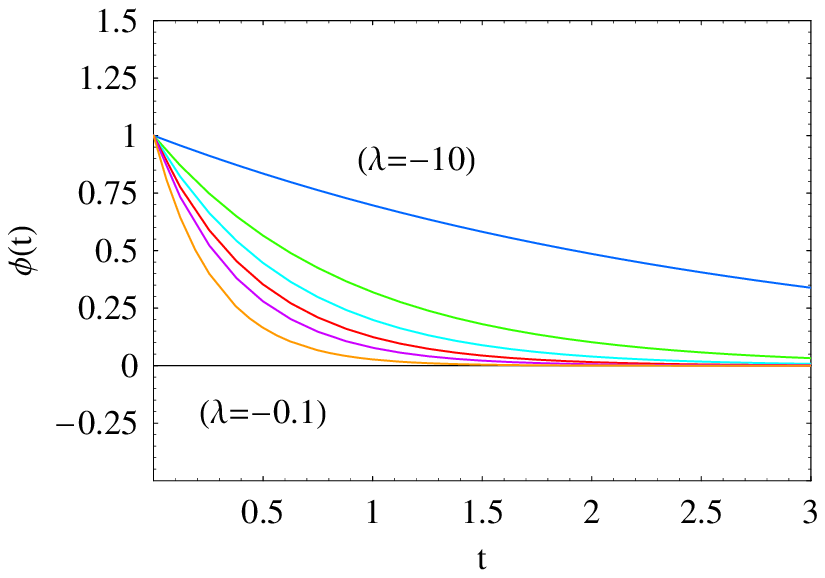}
    \caption{The scale factor $a(t)$ and scalar field $\phi(t)$ versus time for the cosmological
    solutions with $C_1=0$ and $|\lambda|=0.1, 0.2, 0.5, 1, 2, 10$. For simplicity, the
initial values $a_0$ and $\phi_0$ have been normalised to unity.}
   \label{inflation}
\end{figure}

If $C_1=0$, then from Eq. (\ref{general}) it is clear that solutions exist only for
$\lambda <0$. In this case, a simple integration yields the solution: 
\beq
a(t)=a_0\,\exp\biggl(\pm\sqrt{\frac{5}{24 |\lambda|}}\,t\biggr).
\label{sol-C10}
\eeq
The theory admits both increasing and decreasing solutions for the scale factor
with respect to time - choosing the positive sign, we obtain an exponentially
expanding universe with no singularities at finite values of the time coordinate.
Note that no self-coupling potential $V(\phi)$ needed to be introduced for the
scalar field or tailored in any ad hoc way in order to obtain inflation. Instead, it
is the coupling of the scalar field to the quadratic Gauss-Bonnet term that
provides a potential in the most natural way and supports inflationary solutions
in the early universe. Such couplings arise naturally both in the context of
string-inspired or gravitationally modified theories, and we anticipate that
they should all contain similar inflationary solutions - by insisting, however,
in keeping the Ricci scalar in the analysis, these solutions were missed.

The form of the scalar field, in turn, may be easily found via Eq. (\ref{phi-dot}),
that now takes the form
\beq
\frac{\dot \phi}{\phi}=12\lambda\,\frac{\dot a^3}{a^3}\,.
\label{dotphi-dota}
\eeq
Employing the solution for the scale factor (\ref{sol-C10}) found above (with the
positive sign), the scalar field is found to have the form
\beq
\phi=\phi_0\,\exp\biggl(-\frac{5}{4}\,\sqrt{\frac{5}{6|\lambda|}}\,t\biggr).
\label{phi-deSitter}
\eeq
The scalar field is also everywhere well-defined and is decaying exponentially
from an arbitrary initial value $\phi_0$ to zero. The profiles of both the scale
factor and the scalar field are shown in Figs. \ref{inflation}(a,b), respectively, for the
values $|\lambda|=0.1,0.2,0.5,1,2$ and 10; we observe that
the smaller the value of the coupling constant $\lambda$, the faster both
quantities evolve with time.

The effective potential of the scalar field receives contributions from both the
coupling function and the Gauss-Bonnet term, and has the form 
$V_{eff}=-f(\phi) R^2_{GB}/8 \sim \phi^2/|\lambda|$. 
Further aspects of this de Sitter solution were studied in a previous short work
\cite{KGD-short}. In there, we showed that the necessary number of e-foldings
follow easily without the need for assuming trans-planckian initial values $\phi_0$
for the scalar field.  As a result, the effective potential, although of a similar form
to that of chaotic inflation \cite{Linde}, remains always bounded. Its dependence
on the coupling constant $\lambda$ allows also for its value, at the time of
inflation, to be large enough so that it dominates indeed over the other 
constituents of the universe. Finally, its quadratic form places it in the group
of inflationary models that are still compatible with the current observational constraints
\cite{Planck}.


\subsection{The case with $C_1>0$ and $\lambda<0$}

\begin{figure}
\includegraphics[width = 0.48 \textwidth] {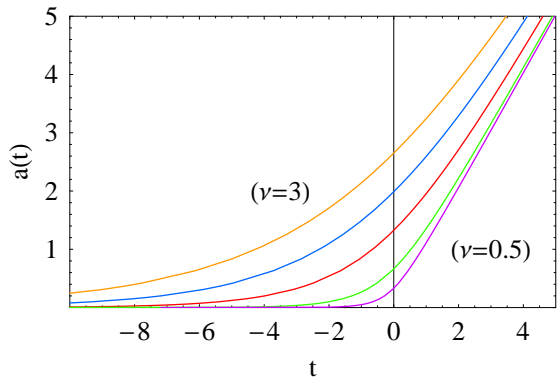}
    \caption{Cosmological solutions with $C_1>0$ and $\lambda<0$, and for
	$\nu=0.5, 0.8,1,2,3$.}
   \label{Milne}
\end{figure}

We now assume that the coupling parameter $\lambda$ is again negative as before,
but we allow the integration constant $C_1$ to take positive values. This class of solutions
first appeared in \cite{KGD-short} but we include it also in the analysis here for 
completeness and for providing additional mathematical details left out in the previous
work. In this case, Eq. (\ref{general}) can be rewritten as:
\beq
\int\,\frac{da}{a}\,\sqrt{a^2+\nu^2} = \pm \sqrt{\frac{5}{2 C_1}}\,(t+t_0)\,,
\eeq
where $\nu^2 \equiv 12 |\lambda|/C_1$. By setting $a=\nu\,\tan w$ and using standard
techniques of integration, the above equation eventually leads to the result
\beq
\sqrt{a^2+\nu^2} + \nu \ln\left(\frac{\sqrt{a^2+ \nu^2}-\nu}{a} \right)
= \pm \sqrt{\frac{5}{2 C_1}}\,(t+t_0)\,.
\label{sol-lam-neg}
\eeq
For $\nu=0$, i.e. in the absence of the Gauss-Bonnet term, we smoothly recover
the linearly expanding solution found in Section 3. For $\nu \neq 0$, Eq. (\ref{sol-lam-neg})
is of a transcendental form and thus cannot be solved for $a$ as a function of $t$.
Its behaviour in terms of time is nevertheless depicted in Fig. \ref{Milne}: 
choosing the (+)-sign, we find that the scale factor is
expanding with time, first with a much faster pace and later with a significantly
slower one. As $\nu$ increases, the point where the scale factor vanishes, i.e. the
initial singularity, moves gradually towards larger negative values of the time
coordinate; on the other hand, for larger values of time, the variation of $\nu$
affects much less the rate of expansion. We may derive analytic asymptotic solutions
describing the behaviour of the scale factor at early times and later times, by
taking appropriate limits of the complete solution (\ref{sol-lam-neg}). To this
end, we first consider the limit $a \rightarrow 0$, in which case we recover the
pure de Sitter solution (\ref{sol-C10}) found in the previous subsection; we thus
conclude that the families of solutions with $\lambda<0$ and with $C_1=0$ and
$C_1>0$ are smoothly connected in the phase-space of the theory. On the other hand,
expanding for large values of $a$, i.e. for $a^2 \gg \nu^2$, we obtain a linear 
function of time for the scale factor. Therefore, in this class of solutions, the
universe interpolates between a pure de Sitter solution at very early times, 
triggered by the Gauss-Bonnet term, and a linearly-expanding Milne-type phase
at later times when the effect of the Gauss-Bonnet term starts to wear off. 
In this way, these solutions may readily accommodate an early inflationary
phase with a natural exit mechanism at later times.

The solution for the scalar field should in principle follow from Eq. (\ref{phi-dot}), however
the absence of an explicit solution for the scale factor in terms of time complicates the
necessary integration. However, we may find an implicit expression for the scalar field in
the following way: if we use Eq. (\ref{phi-dot}), then Eq. (\ref{Eq1and2-v2}) may be
re-written, for a general polynomial coupling function, as
\beq
-2(n-1)\frac{\dot \phi}{\phi}=\frac{\ddot a\,(k+5 {\dot a}^2)}
{\dot a\,(k+{\dot a}^2)}\,,
\eeq
which, upon integration of both sides with respect to time, gives the constraint
\beq
\phi^{2(n-1)}=\frac{C_0}{\dot a\,(k+{\dot a}^2)^2}\,.
\label{adotphi}
\eeq
Specialising for the case $k=0$ and $n=2$, and using Eq. (\ref{general}) to replace
$\dot a$, we finally obtain
\beq
\phi^2=C_0\,\left(\frac{2C_1}{5}\right)^{5/2}\frac{(a^2+\nu^2)^{5/2}}{a^5}\,.
\eeq
Using the asymptotic behaviour of the scale factor at early times, i.e. the pure
de Sitter solution (\ref{sol-C10}), we may easily conclude that the scalar field
also assumes its exponentially decaying form of Eq. (\ref{phi-deSitter}). On the
other hand, for $a^2 \gg \nu^2$, the scalar field reduces to a constant. 


\subsection{The case with $C_1<0$ and $\lambda<0$}

\begin{figure}
\includegraphics[width = 0.48 \textwidth] {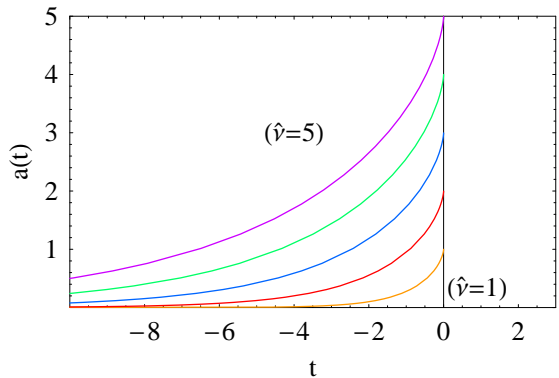}
    \caption{Cosmological solutions with $C_1<0$ and $\lambda<0$, and for
	$\hat \nu=1,2,3,4,5$ from left to right.}
   \label{C1neg_lamneg}
\end{figure}

We will next assume that $\lambda<0$ again, but that $C_1<0$ now. From Eq. (\ref{general}),
we may see that any solutions that would follow in this case will not allow the scale
factor to grow indefinitely but only up to a maximum value, otherwise the rate of
expansion would become imaginary. Equation (\ref{general}), for this choice of
parameters, can be alternatively written as:
\beq
\int\,\frac{da}{a}\,\sqrt{\hat \nu^2-a^2} = \pm \sqrt{\frac{5}{2 |C_1|}}\,(t+t_0)\,,
\eeq
where $\hat \nu^2 \equiv 12 |\lambda|/|C_1|$. We now set $a=\hat\nu\,\sin w$, and
by integrating once we find the result
\beq
\sqrt{\hat \nu^2 -a^2} +\hat \nu \ln\left(\frac{\hat \nu -\sqrt{\hat \nu^2 -a^2}}{a} \right)
= \pm \sqrt{\frac{5}{2 |C_1|}}\,(t+t_0)\,.
\label{sol-lam-pos-2}
\eeq
The profile of the scale factor in terms of time, as this follows from the above
relation for the positive sign, is depicted in Fig. \ref{C1neg_lamneg}. 
The early-time behaviour of the above solution is very similar to the one found in the
previous subsection: a singularity, where $a=0$, is again present but this is reached 
at increasingly large negative values of the time coordinate as $\hat \nu$ increases;
the expansion of the relation (\ref{sol-lam-pos-2}) in the limit $a \rightarrow 0$
leads once again to the pure de Sitter solution (\ref{sol-C10}) and to an inflationary
phase for the universe. However, the quantity $\hat \nu$ is now the maximum allowed
value of $a(t)$, and the universe stops increasing after this point. We thus conclude
that, in the phase space of the solutions of the theory, the pure de Sitter solutions
for $\lambda<0$ and $C_1=0$ branch off to two families of cosmological solutions
with totally different behaviour at larger times: one that allows for indefinite expansion
of the universe for $C_1>0$ and one where the scale factor reaches a ceiling 
for $C_1<0$. 

The solution for the scalar field follows again from Eq. (\ref{adotphi}) and in this
case is given by
\beq
\phi^2=C_0\,\left(\frac{2|C_1|}{5}\right)^{5/2}\frac{(\hat \nu^2-a^2)^{5/2}}{a^5}\,.
\eeq
Again, in the early-time limit where $a \rightarrow 0$, the scalar field adopts the
exponentially decaying form of Eq. (\ref{phi-deSitter}). On the other hand,
for $a^2  \rightarrow \hat \nu^2$, the scalar field goes to zero. As in the
previous case, the scalar field remains finite and diverges only at the 
initial singularity where the scale factor vanishes.


\subsection{The case with $C_1>0$ and $\lambda>0$}

\begin{figure}
\includegraphics[width = 0.48 \textwidth] {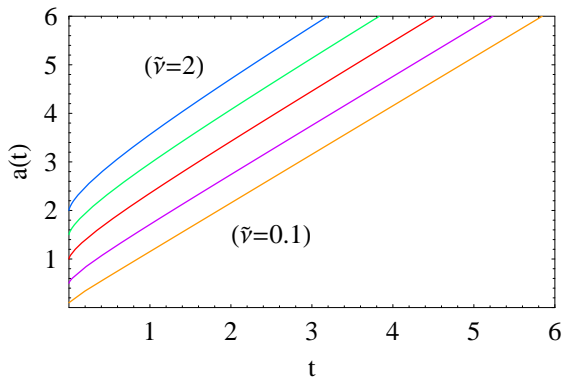}
    \caption{Cosmological solutions with $C_1>0$ and $\lambda>0$, and for
	$\tilde \nu=0.1,0.5,1,1.5,2$ from left to right.}
   \label{C1pos_lampos}
\end{figure}

We now reverse the sign of the coupling parameter $\lambda$ and allow it to take
only positive values. In that case, the choices $C_1 \leq 0$ are not allowed, and
mathematically consistent solutions may be derived only for $C_1>0$. In this case,
Eq. (\ref{general}) can be rewritten as:
\beq
\int\,\frac{da}{a}\,\sqrt{a^2-\tilde \nu^2} = \pm \sqrt{\frac{5}{2 C_1}}\,(t+t_0)\,,
\eeq
where $\tilde \nu^2 \equiv 12 \lambda/C_1$. We now set $a=\tilde \nu/\cos w$, 
and upon integrating, we find the relation 
\beq
\sqrt{a^2-\tilde \nu^2} -\tilde \nu \arccos\left(\frac{\tilde \nu}{a}\right)=
\pm \sqrt{\frac{5}{2 C_1}}\,(t+t_0)\,.
\label{sol-lam-pos}
\eeq
For $\tilde \nu=0$, i.e. $\lambda=0$, we may easily see that we go back to a linearly
expanding solution with an initial singularity at a finite value of the time coordinate.
However, for $\tilde\nu \neq 0$, i.e. in the presence of the Gauss-Bonnet term, the
situation is radically different: the square-root on the left-hand-side demands that
$a^2 \geq \tilde \nu^2$, therefore $\tilde \nu$ becomes the smallest allowed value
of the scale factor in this model, and no singularities are allowed to arise. The same
conclusion follows from the inverse cosine function whose argument should also satisfy
the inequality $-1 \leq \tilde \nu/a \leq 1$, or  $\tilde \nu \leq a$ in our case since both
$\tilde \nu$ and $a$ are positive-definite.
If we expand the solution (\ref{sol-lam-pos}) for values of $a$ close to its
minimum value $\tilde \nu$, we find the asymptotic solution
\beq
a(t) \simeq \tilde \nu\,[1+ (A t+B)^{2/3}]\,,
\eeq
which shows its regular behaviour for any finite values of the time-coordinate.
In Fig. \ref{C1pos_lampos}, we plot the behaviour of the scale factor as a function of time: the
different curves correspond to different values of $\tilde \nu=0.1,0.5,1,1.5,2$ from left
to right. We observe that as $\tilde \nu$ increases the curve is moving towards larger
values of $a$ and thus away from the singularities. For large values of $a$, the
dependence becomes linear again in terms of time -- note that this behaviour is
common for the two solutions arising for $C_1>0$. 

The implicit solution for the scalar field in this case is given by
\beq
\phi^2=C_0\,\left(\frac{2C_1}{5}\right)^{5/2}\frac{(a^2-\tilde \nu^2)^{5/2}}{a^5}\,.
\eeq
Since the minimum value of the scale factor is $\tilde \nu$, the above expression
for the scalar field remains always finite. It starts from a zero value at very
early time, it follows an increasing profile reaching eventually a constant value.

Let us comment at this point on the implications of the existence of this particular
class of solutions. Singularity-free solutions arising in the context of string-inspired
theories, or modified gravitational theories in general, are always important as they
provide an alternative to the cosmological solutions of the classical theory of
General Relativity that always possess an initial (or final) singularity. Solutions
free from the initial singularity have been derived, for instance, in the context of the
heterotic superstring effective theory \cite{Antoniadis}. In there, the coupling function between
the scalar field and the Gauss-Bonnet term is given by the highly non-trivial
expression $f(\phi)=\ln [2 e^\phi \eta(i e^\phi)]$, where $\eta$ is the Dedekind
function. However, in \cite{KRT} analytical arguments were presented that showed
that singularity-free solutions may arise in the context of a similar theory where the
coupling function has a much simpler form, namely $f(\phi)=\lambda \phi^n$,
where $n$ was strictly an even, positive number -- specific solutions were found numerically
in the context of the same work. The reason was that these two different
forms share in fact three important characteristics: they are invariant under the change
$\phi \rightarrow -\phi$, they have a global minimum and asymptotically they
tend to infinity. 

In the present work, we have indeed managed to derive an 
analytical singularity-free solution for the even coupling function $f(\phi)=\lambda \phi^2$,
in total agreement with the aforementioned argument. Note, however, that the analysis
of \cite{KRT} took into account the presence of the Ricci scalar in the theory thus forcing
the authors to perform numerical integration in order to find the sought solutions.
Here, by ignoring the Ricci scalar, we have managed to demonstrate in a very simple
way the emergence of singularity-free solutions and  to find their exact analytical form.
We believe that this result solidifies even more our argument that the Ricci scalar may
indeed be ignored in the early-time limit with no effect in the dynamics of the universe
and that, in fact, one should do so, in order to derive elegant, analytical solutions.

%

\section{The Case of a General Polynomial Coupling Function}

In this section, we will briefly address the case of a general polynomial coupling
function, $f(\phi)=\lambda \phi^n$. Both Eqs. (\ref{Eq1and2-v2}) and (\ref{adotphi})
have been derived for this case, and thus can be straightforwardly used for our
purpose. Solving Eq. (\ref{adotphi}) for the scalar field $\phi$ and substituting
in Eq. (\ref{Eq1and2-v2}), we obtain the constraint
\begin{align}
(k+5\dot a^2)\,\ddot a \,\dot a^{\frac{2-3n}{2 (n-1)}}a^3+12\Gamma (k+\dot a^2)^{n/(n-1)}=0\,,
\end{align}
with
\beq
\Gamma \equiv \lambda n (n-1)\, C_0^{\frac{n-2}{n-1}}\,.
\label{Gamma-def}
\eeq

Specializing to the case of a flat universe with $k=0$, we obtain
\begin{align}
5\ddot a \,\dot a^{\frac{3n+2}{2 (1-n)}}a^3+12\Gamma =0\,.
\label{eq:n}
\end{align}
We can easily check that the above equation admits de Sitter-type solutions only for
particular values of the integer $n$. Setting $a(t)=\exp(H_0\,t)$, the aforementioned equation
becomes
\begin{align}
5 H_0^{\frac{n-6}{2 (n-1)}}\exp\left[\frac{5}{2}\,\frac{(n-2)}{(n-1)}\,H_0\,t\right]+12 \Gamma=0\,.
\end{align}
According to the above, consistent de Sitter-type solutions arise only for $n=2$, and for
$\Gamma<0$ or equivalently $\lambda<0$, in accordance to the results of Section 5. 

If we integrate Eq. (\ref{eq:n}) with respect to time once, we find the relation
\begin{align}
\dot a^p=\frac{\alpha}{a^2}+\beta\,,
\label{ap}
\end{align}
where we have defined the parameters
\beq
\alpha \equiv \frac{3 (n-6)}{5(n-1)}\,\Gamma\,, \qquad ~~p=\frac{n-6}{2 (n-1)}\,,
\label{alpha-p}
\eeq
and $\beta$ is an arbitrary integration parameter.
We observe that for large values of the scale factor, the term $\alpha/a^2$ on the
right-hand-side of Eq. (\ref{ap}) can be ignored. Then, for $\beta=0$, the universe
reaches asymptotically a static Einstein-type solution, while for $\beta \neq 0$, we obtain
a universal behaviour of the form $a(t)=A t+ B$ independently of the exact value of $n$.

\begin{figure}
\includegraphics[width = 0.48 \textwidth] {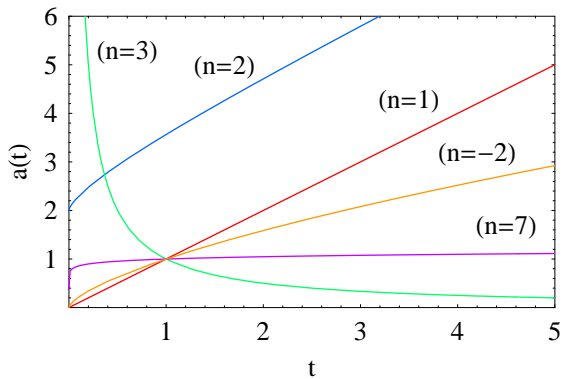}
    \caption{Early-time cosmological solutions in the scalar-GB theory with a general
polynomial coupling function $f(\phi)=\lambda \phi^n$ and for various values of $n$.}
   \label{General-n}
\end{figure}

On the other hand, in the early universe where $a$ is small, we can approximate 
Eq. (\ref{ap}) by the following equation
\begin{align}
\dot a^p=\frac{\alpha}{a^2}\,,
\end{align}
that upon integration, and for $\alpha>0$ or equivalently $\lambda>0$, gives
\begin{align}
a(t)=\Bigl(At+B\Bigr)^{\frac{n-6}{5(n-2)}}\,. \label{a-early}
\end{align}
Contrary to the universal behaviour characterising the large-$a$ limit found above,
the dynamics of the universe in the small-$a$ limit, i.e. in the strong-curvature
regime, strongly depends on the exact form of the coupling function and thus on the
value of the integer $n$. Note that due to the expression of the solution (\ref{a-early}),
the case with $n=2$ -- that was studied in detail in Section 5 - should be excluded;
therefore, the solution (\ref{a-early}) describes the early-universe dynamics for solutions
obtained in the context of the scalar-GB theory with a polynomial coupling function with $n \neq 2$.
For instance, for $n=1$, we correctly recover the linear early-time solution $a(t)=At + B$
obtained in Section 4. In general, all cases with $n<2$ or $n>6$ lead to expanding cosmological
solutions with an initial singularity at a finite value of the time-coordinate, whereas all
cases with $2<n<6$ describe contracting cosmological solutions with a final singularity
at asymptotically infinite time. Indicative cases of the above solutions are depicted in
Fig. \ref{General-n}. Clearly, it is only for the case with $n=2$ and $\lambda>0$ -- depicted also in 
Fig. \ref{General-n} for completeness -- that an expanding singularity-free solution arises
in the theory.

%

\section{A Transitory Solution}

Looking at the gravitational equations (\ref{Ein-1})-(\ref{Ein-2}), we conclude that, by
ignoring the Ricci scalar in the previous sections, what we have actually assumed is
the following conditions to hold
\beq
\dot f\,\frac{\dot a}{a} \gg 1\,, \qquad\qquad  \ddot f \gg 1\,,
\label{conditions}
\eeq
even for a general coupling function $f(\phi)$. In sections 3 and 4, we demonstrated,
for two different choices of the coupling function, that the Ricci scalar can indeed be
ignored in the early-time limit and therefore the above conditions hold.

In order to study how the system will transit from the very early-time epoch, where the
Gauss-Bonnet term dominates, to a subsequent era, where the Ricci scalar starts becoming
important, here we will assume that the following constraint holds between $f$ and the scale
factor
\beq
\dot f\,\frac{\dot a}{a}=c_0\,.
\label{constraint}
\eeq
In the above, $c_0$ is an arbitrary constant. 
We will thus assume that there is an era where the Gauss-Bonnet
term starts giving a contribution to the gravitational field equations that is of the same order
as that of the Ricci scalar. We will first consider the case where $c_0$ is still much larger than unity, 
therefore, Ricci scalar can still be ignored. Then, we will assume that $c_0 \simeq 1$,
in which case both gravitational terms should be taken into account in the field equations.
Finally, the limit $c_0 \ll 1$ will be considered, valid in a era where the Ricci scalar will be
the dominant term. Although the exact study of the evolution of the system demands a
numerical analysis, we believe that the following study will give us a feeling of whether
such a transition is possible.

Assuming that the constraint (\ref{constraint}) holds and that $c_0 \gg 1$, the Ricci scalar
may be ignored and the corresponding field equation (\ref{Ein-1-new}) is re-written as
\beq
\dot \phi^2=6c_0\,\frac{(k+\dot a^2)}{a^2}\,.
\label{phi-dot-special-1}
\eeq
Also, differentiating Eq. (\ref{constraint}) once with respect to time, we obtain
\beq
\ddot f=\dot f\,\biggl(\frac{\dot a}{a}-\frac{\ddot a}{\dot a}\biggr)\,.
\label{ddotf-special}
\eeq
When the above and Eq. (\ref{phi-dot-special-1}) are used into the second gravitational
equation (\ref{Ein-2-new}), the latter also takes the new form
\beq
\frac{4\dot a}{a}=\frac{\ddot a}{\dot a}\,
\frac{(k -\dot a^2)}{(k+\dot a^2)}\,.
\eeq
The above can easily be integrated once, using elementary methods, and
leads to the relation
\beq
(k+\dot a^2)\,a^4=C\,\dot a\,.
\eeq
Rearranging the above and integrating once more, we obtain the integral equation
\beq
\int\,\frac{2 a^4 da}{C + \epsilon \sqrt{C^2-4ka^8}}=t+t_0\,,
\label{special-k}
\eeq
where $\epsilon = \pm 1$. For $k \neq 0$, this arbitrary sign denotes the presence of two
distinct branch solutions that have the form
\beq
\frac{8\epsilon a^5}{15C}\,F\left(\frac{1}{2},\frac{5}{8},\frac{13}{8};\frac{4ka^8}{C^2}\right)
+\frac{\epsilon \sqrt{C^2-4ka^8}-C}{6ka^3}=t+t_0\,,
\label{solution-kplusbr}
\eeq
where $F(a,b,c;x)$ is again the hypergeometric function. Above, we have assumed that the
integration constant $C$ is positive -- in fact, there is a symmetry between the signs of
$C$ and $\epsilon$, therefore we are allowed to fix the sign of one of the two parameters.
Depending on the values of $k= \pm$ and $\epsilon=\pm 1$, the above expression describes
a variety of smooth cosmological solutions with various asymptotic behaviours as
$a \rightarrow 0$ or $a \rightarrow \infty$. For instance, for $\epsilon=-1$ and for
both values of $k=\pm 1$, solutions arise that do not possess any singularity at a 
finite value of the time-coordinate. 

However, here we are mainly interested in studying the transition of the system through the
different epochs, and for this reason we may simplify our analysis by setting $k=0$. 
Then, from Eq. (\ref{special-k}), we must necessarily have $\epsilon=1$, and a simple
integration leads to the power-law solution
\beq
a(t)=(At+B)^{1/5}\,, \label{scale-15}
\eeq
where $A$ and $B$ are again integration constants. The solution for the scalar field may
easily follow from Eq. (\ref{phi-dot-special-1}) upon using the above solution for the scale factor;
it has the form
\beq
\phi(t)=\phi_0 \pm \frac{\sqrt{6c_0}}{5}\,\ln(At+B)\,.
\eeq
For these simple forms of the scale factor and scalar field, one may determine the
form of the coupling function $f(\phi)$: integrating Eq. (\ref{constraint}), we obtain
\beq
f(t)=f_0 +\frac{5c_0}{2A^2}\,(At +B)^2\,,
\eeq
which may be alternatively written as
\beq
f(\phi) =f_0+f_1\,\exp\left(\pm\frac{10}{\sqrt{6c_0}}\,\phi\right)\,.
\eeq
Therefore, the above special solution actually corresponds to a particular
choice of an exponential coupling function.

\begin{figure}
\includegraphics[width = 0.48 \textwidth] {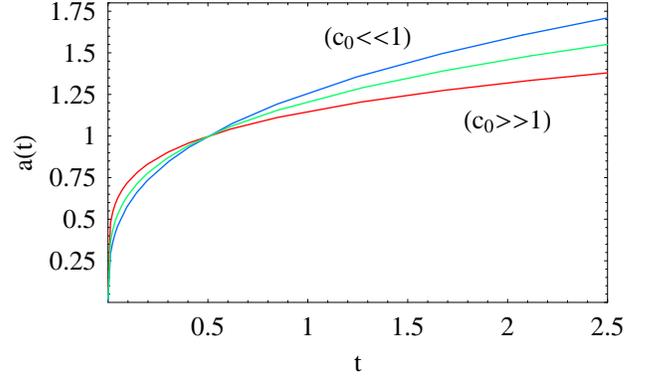}
    \caption{An indicative transitory solution with the value of the parameter $c_0$
    parametrising the weight of the GB term in the theory -- the curve in the middle
corresponds to the limiting case with $c_0 \simeq 1$.}
   \label{Transitory}
\end{figure}

We now assume that, as the time goes by, the effect of the Gauss-Bonnet term begins
to diminish, and the point is reached where $c_0 \simeq 1$. Then, the contribution
of the Ricci scalar must be restored, and in that case Eq. (\ref{Ein-1})
leads to the result
\beq
\dot \phi^2=6(1+c_0)\,\frac{(k+\dot a^2)}{a^2}\,.
\label{phi-dot-special-2}
\eeq
Then, using both (\ref{phi-dot-special-2}) and (\ref{ddotf-special}) in 
Eq. (\ref{Ein-2}), we take the alternative constraint
\beq
4(1+c_0)\,\frac{(k+\dot a^2)}{a^2} + (1+c_0)\,\frac{2\ddot a}{a} -
c_0\,\frac{\ddot a}{a}\,\frac{(k+\dot a^2)}{\dot a^2}=0\,,
\eeq
or, after a little bit of algebra,
\beq
4(1+c_0)\,\frac{\dot a}{a}=\frac{\ddot a}{\dot a}\,
\frac{[c_0 k -(2+c_0)\dot a^2]}{(k+\dot a^2)}\,.
\eeq
The above can be integrated once with respect to time to give the relation
\beq
(k+\dot a^2)\,a^4=C\,\dot a^{c_0/(1+c_0)}\,,
\eeq
where $C$ an integration constant. Once again, for $k \neq 0$, the analytical calculation
is extremely difficult as the above equation exhibits a non-algebraic dependence. Therefore,
we set $k=0$ and, upon integrating once more, we find the solution
\beq
a(t)=(A t + B)^{(2+c_0)/(6+5 c_0)}\,.
\eeq
From Eq. (\ref{phi-dot-special-2}), we see that, for $k=0$, $1+c_0>0$, therefore the
power of $(At +B)$ in the above expression is also positive. Therefore, the above describes
an expanding universe with an initial singularity emerging when $A t+B \rightarrow 0$
 i.e. at a finite time-coordinate. For $c_0 \gg 1$, the above
power reduces to 1/5, as expected, in accordance with Eq. (\ref{scale-15}). In the
limit $c_0 \ll 1$, the power becomes 1/3. For $c_0 \simeq 1$, the power interpolates
between 1/5 and 1/3. In Fig. \ref{Transitory}, we plot the two indicative cases with
$c_0 \gg 1$ and $c_0 \ll 1$ as well as the limiting case with $c_0 \simeq 1$ that lies in
between. Although the initial singularity in this particular solution cannot
be altogether avoided, the presence of the Gauss-Bonnet term works towards making the
singularity softer. As the universe expands, the rate of expansion increases due to the decrease
of the effect of the Gauss-Bonnet term reaching eventually its highest value, taken in
the context of the pure Einstein-scalar field theory, when the Gauss-Bonnet becomes
negligible. Of course, a complete cosmological analysis would demand the introduction
of additional ingredients in the universe, such as the radiation or matter energy density,
and the detailed study of the sequence of different eras - however, this is beyond
the scope of the present work that investigates the role of the Gauss-Bonnet term
in the very-early-universe dynamics.


\section{Conclusions}

In this work, we have considered a generalised gravitational theory that contained,
apart from the Einstein term, a scalar field and a higher-curvature
-- the quadratic -- Gauss-Bonnet term. Both of these additions are met in the
superstring effective theory as well as in a variety of modified gravitational theories
considered in the literature over the years. Theories of this type, with either 
stringy-like or more general coupling functions between the scalar field and the 
Gauss-Bonnet term -- a necessary feature in order for the GB term to remain in the
theory -- have shown to lead to novel gravitational solutions. Here, we have focused
on the early-universe dynamics and demonstrated that a simple choice of the coupling
function and a simplifying assumption regarding the role of the Ricci scalar can lead
to new, analytical, elegant solutions with interesting characteristics.

Starting from the latter basic element of our analysis -- the role of the Ricci
scalar in the early-universe dynamics, a toy model, that was considered in Section 3,
hinted to the fact that the presence of the Ricci scalar in the theory merely makes
the analysis much more complex without affecting the actual solutions for the
scale factor of the universe and the scalar field. This hint was changed to a
certainty when, in Section 4, the Einstein-scalar-GB theory with a linear coupling
function was considered, and
the early-time limit of the complete solution of the set of field equations was
derived; it was found to be identical to the solution derived in the context of
the pure scalar-GB theory. As expected, the higher-curvature -- quadratic --
GB term dominates over the linear Ricci scalar and, in conjunction to the scalar
field, determines the form of the cosmological solution at early times when the
curvature is strong. That is, in the very early universe it would be higher-order
curvature terms that would be dominant; therefore, we should have a theory involving
higher orders of Riemann curvature, and yet one that should be ghost-free. This picks
out pure Lovelock theory, which is a homogeneous polynomial of degree $N$ in Riemann
tensor where linear $N=1$ is Einstein and quadratic $N=2$ is GB. 

Guided by the aforementioned results, in Section 5 we proceeded to study the
pure scalar-GB theory this time with a quadratic coupling function between the scalar 
field and the GB term. Although of a simple form, this choice led to a plethora of
solutions with interesting characteristics: for a negative coupling parameter, the
set of equations supported solutions that were either inflationary, de Sitter-type
or more involved expanding solutions with a de Sitter phase in their past and a
natural exit mechanism at later times; for a positive coupling function, the set
of equations gave rise to singularity-free solutions with no Big-Bang singularity.
All these solutions were derived in an analytical, elegant way that allowed for
the comprehensive study of their properties instead of the numerical study that
is usually employed in the context of similar generalised gravitational theories.
  
The case of the general polynomial coupling function was briefly considered in
Section 6. There, it was demonstrated that inflationary, de Sitter-type solutions 
arise indeed only for the case of a quadratic coupling function and for no other.
The asymptotic behaviour of the scale factor of the universe was derived, and
shown that for large values of $a$, the universe adopts a universal behaviour
independently of the exact form of the polynomial coupling function; for small
values of $a$, on the other hand, the form of the scale factor depends strongly
on the value of the integer $n$, that determines the power of the polynomial
function: again, in the strong curvature regime, it is only the case of $n=2$
that leads to singularity-free solutions whereas all the other choices lead to
solutions with either an initial or a final singularity.

Naturally, as the universe expands, its curvature becomes smaller and the role 
of the Ricci scalar will gradually start being of importance again. A toy transitory
solution was considered in Section 7, that helps to visualise how the transitions
between the different eras take place:  starting from the era where the Ricci scalar
can be totally ignored, passing to the era where it should be taken into
account and finally arriving at the one where the Ricci scalar dominates thus
restoring the traditional cosmology. 

Our analysis has singled out the quadratic coupling function of the scalar field
to the Gauss-Bonnet term as one of particular importance: it is only for this
choice that inflationary solutions as well as singularity-free solutions -- two
classes of solutions of significant interest in cosmology -- emerge in the context
of the theory. We do not consider this to be accidental for the following two reasons: 
on the one hand, the quadratic inflaton potential is the only polynomial form
that is still compatible with the current observational constraints \cite{Planck}; 
in our inflationary solutions, the inflaton effective potential is again quadratic
and, although similar to the one of chaotic inflation, it avoids the caveat of the
trans-planckian initial conditions \cite{KGD-short}.
It is worth noting that this attractive inflationary model follows from a legitimate
string-inspired theory rather than being built in an artificial and ad hoc way
as is usually seen in the literature. On the other hand, our quadratic coupling
function belongs to a more general class of coupling functions that share a number
of common characteristics with the exact coupling function of the heterotic string 
effective theory. It was in the context of the latter theory that singularity-free solutions
first emerged and the same achievement was performed also in the present analysis. 
We are thus led to conclude that a more fundamental, underlying connection may exist
between the quadratic coupling function and the one of the ultimate theory, a connection
that certainly needs to be investigated further.

\section{Acknowledgments}

This research has been co-financed by the European
Union (European Social Fund - ESF) and Greek national funds through the
Operational Program ``Education and Lifelong Learning'' of the National
Strategic Reference Framework (NSRF) - Research Funding Program: 
``THALIS. Investing in the society of knowledge through the European
Social Fund''. Part of this work was supported by the COST Action MP1210
``The String Theory Universe''.



\begin{thebibliography}{99}

\bibitem{strings} M.~B.~Green, J.~H.~Schwarz and E.~Witten,
{\it ``Superstring Theory''}, Cambridge Monogr. Math. Phys. (1987).

\bibitem{Lovelock} D. Lovelock, J. Math. Phys. {\bf 12}, 498 (1971). 

\bibitem{ADD}
N.~Arkani-Hamed, S.~Dimopoulos and G.~R.~Dvali,
{Phys.\ Lett.}\ B {\bf 429}, 263 (1998) [hep-ph/9803315];
{Phys.\ Rev.}\ D {\bf 59}, 086004 (1999) [hep-ph/9807344];
\\
I.~Antoniadis, N.~Arkani-Hamed, S.~Dimopoulos and G.~R.~Dvali,
{Phys.\ Lett.}\ B {\bf 436}, 257 (1998) [hep-ph/9804398].

\bibitem{RS} L. Randall and R. Sundrum,
{Phys. Rev. Lett.} {\bf 83} (1999) 3370; 
{Phys. Rev. Lett.} {\bf 83} (1999) 4690. 

\bibitem{Zwiebach}
  B.~Zwiebach,
  Phys.\ Lett.\ B {\bf 156} (1985) 315.

\bibitem{Gross} D.~J.~Gross and J.~H.~Sloan,
  Nucl.\ Phys.\ B {\bf 291}, 41 (1987).

\bibitem{Tseytlin} R.~R.~Metsaev and A.~A.~Tseytlin,
  Nucl.\ Phys.\ B {\bf 293}, 385 (1987).

\bibitem{Dadhich} N. Dadhich, Pramana 74, 875 (2010); ``The gravitational equation in higher
dimensions'', in {\it Relativity and Gravitation: 100 years after Einstein in Prague}, eds J. Bicak and T. Ledvinka, Springer (2013) [arxiv:1210.3022], June 25-28, 2012.


\bibitem{Antoniadis} I.~Antoniadis, J.~Rizos and K.~Tamvakis,
  Nucl.\ Phys.\ B {\bf 415}, 497 (1994).

\bibitem{KRT} P.~Kanti, J.~Rizos and K.~Tamvakis,
  Phys.\ Rev.\ D {\bf 59}, 083512 (1999).

\bibitem{KMRTW} P.~Kanti, N.~E.~Mavromatos, J.~Rizos, K.~Tamvakis and E.~Winstanley,
  Phys.\ Rev.\ D {\bf 54}, 5049 (1996); D {\bf 57}, 6255 (1998).

\bibitem{Torii} T.~Torii, H.~Yajima and K.~i.~Maeda,
  Phys.\ Rev.\ D {\bf 55}, 739 (1997).

\bibitem{KKK}  P.~Kanti, B.~Kleihaus and J.~Kunz,
  Phys.\ Rev.\ Lett.\  {\bf 107}, 271101 (2011); Phys.\ Rev.\ D {\bf 85}, 044007 (2012).

\bibitem{Nojiri} S.~Nojiri and S.~D.~Odintsov,
  eConf C {\bf 0602061}, 06 (2006)
  [Int.\ J.\ Geom.\ Meth.\ Mod.\ Phys.\  {\bf 4}, 115 (2007)]
  [hep-th/0601213].



\bibitem{Gregory:2008bf}
  R.~Gregory,
  Prog.\ Theor.\ Phys.\ Suppl.\  {\bf 172} (2008) 71
  [arXiv:0801.1603 [hep-th]].

\bibitem{Chen:2012au}
  T.~j.~Chen, M.~Fasiello, E.~A.~Lim and A.~J.~Tolley,
  JCAP {\bf 1302} (2013) 042
  [arXiv:1209.0583 [hep-th]].

\bibitem{Gleyzes:2014dya}
  J.~Gleyzes, D.~Langlois, F.~Piazza and F.~Vernizzi,
  Phys.\ Rev.\ Lett.\  {\bf 114} (2015) 21,  211101
  [arXiv:1404.6495 [hep-th]].


\bibitem{Amendola:2007ni}
  L.~Amendola, C.~Charmousis and S.~C.~Davis,
  JCAP {\bf 0710} (2007) 004
  [arXiv:0704.0175 [astro-ph]].

\bibitem{Koivisto:2006xf}
  T.~Koivisto and D.~F.~Mota,
  Phys.\ Lett.\ B {\bf 644} (2007) 104
  [astro-ph/0606078].

\bibitem{Carter:2005fu}
  B.~M.~N.~Carter and I.~P.~Neupane,
  JCAP {\bf 0606} (2006) 004
  [hep-th/0512262].

\bibitem{Leith:2007bu}
  B.~M.~Leith and I.~P.~Neupane,
  JCAP {\bf 0705} (2007) 019
  [hep-th/0702002].


\bibitem{Lifshitz:1963ps}
  E.~M.~Lifshitz and I.~M.~Khalatnikov,
  Adv.\ Phys.\  {\bf 12} (1963) 185.
  
\bibitem{Belinsky:1970ew}
  V.~A.~Belinsky, I.~M.~Khalatnikov and E.~M.~Lifshitz,
  Adv.\ Phys.\  {\bf 19} (1970) 525.

\bibitem{Belinskii:1972sg}
  V.~A.~Belinskii, E.~M.~Lifshitz and I.~M.~Khalatnikov,
  Zh.\ Eksp.\ Teor.\ Fiz.\  {\bf 62} (1972) 1606.

\bibitem{KGD-short}  P.~Kanti, R.~Gannouji and N.~Dadhich,
  arXiv:1503.01579 [hep-th].

\bibitem{Guth} A.~H.~Guth,
  Phys.\ Rev.\ D {\bf 23}, 347 (1981).

\bibitem{Linde} A.D. Linde, Phys. Lett. B {\bf 129}, 177 (1983).

\bibitem{Starobinsky} A.A. Starobinsky,  Phys. Lett. B {\bf 91}, 99 (1980).

\bibitem{Kinney:2005vj} W.~H.~Kinney,
  Phys.\ Rev.\ D {72} (2005) 023515.
  
\bibitem{Motohashi:2014ppa} H.~Motohashi, A.~A.~Starobinsky and J.~Yokoyama,
  arXiv:1411.5021 [astro-ph.CO].

\bibitem{Planck} P.~A.~R.~Ade {\it et al.}  [Planck Collaboration],
  arXiv:1502.02114 [astro-ph.CO].
  
\end{thebibliography}
\end{document}